\documentclass[conference]{IEEEtran}

\usepackage{graphicx}
\usepackage{psfig,epsfig}
\usepackage{multirow}
\usepackage{amsmath}
\usepackage{multicol}
\usepackage{hyperref}
\ifCLASSINFOpdf

\else

\fi

\usepackage{booktabs}
\usepackage{diagbox}

\begin{document}


\title{I-MSV 2022: Indic-Multilingual and Multi-sensor Speaker Verification Challenge}



   \author{\IEEEauthorblockN{ Jagabandhu Mishra, Mrinmoy Bhattacharjee and S. R. Mahadeva Prasanna}
\IEEEauthorblockA{\\Department of Electrical Engineering\\
Indian Institute of Technology Dharwad, Dharwad-580011, India\\
Email: \{jagabandhu.mishra.18, mrinmoy.b, prasanna\}@iitdh.ac.in}}

\maketitle
\begin{abstract}
Speaker Verification (SV) is a task to verify the claimed identity of the claimant using his/her voice sample. Though there exists an ample amount of research in SV technologies, the development concerning a multilingual conversation is limited.  In a country like India, almost all the speakers are polyglot in nature. Consequently, the development of a Multilingual SV (MSV) system on the data collected in the Indian scenario is more challenging. With this motivation, the Indic- Multilingual Speaker Verification (I-MSV) Challenge 2022 has been designed for understanding and comparing the state-of-the-art SV techniques. For the challenge, approximately $100$ hours of data spoken by $100$ speakers has been collected using $5$ different sensors in $13$ Indian languages. The data is divided into development, training, and testing sets and has been made publicly available for further research. The goal of this challenge is to make the SV system robust to language and sensor variations between enrollment and testing. In the challenge, participants were asked to develop the SV  system  in two scenarios, viz. constrained and unconstrained. The best system in the constrained and unconstrained scenario achieved a performance of $2.12\%$ and $0.26\%$ in terms of Equal Error Rate (EER), respectively.
\end{abstract}

\section{Introduction}
Speaker Verification (SV) is the task of validating the identity of a speaker using the voice sample of the claimant. The tremendous development in SV technology in the last five decades has enabled the system to be deployed in various application areas, starting from voice-based attendance system to authentication for bank transactions~\cite{bai2021speaker}. However, the performance of the systems suffer when multiple languages and sensors are involved during testing~\cite{khosravani2017plda}. Hence, the scalability of SV systems is limited considering such scenarios. The citizens of India use approximately $122$ major and $1599$ other languages in their day-to-day conversation. Most importantly, they are polyglot in nature. Therefore, the flexibility in language and sensors during testing may restrict the reach of SV technologies. With this motivation, the Indian Institute of Technology Guwahati Multi Variability (IITG-MV) data was collected using five different sensors from the people coming from different geographical locations of India having variations in the native language, dialect, and accent~\cite{haris2012multivariability}. 

In the literature, there exist few works on the development of SV in multilingual and domain mismatch scenarios~\cite{khosravani2017plda}. The reported works contribute to the feature, model, and score level for minimizing the impact of language and domain mismatch~\cite{khosravani2017plda}. Most of the reported work uses either an in-house dataset or publicly available data (mostly crawled from the public domain) for performing their studies. The in-house data are limited by the number of speakers, languages, and sensors. Though the publicly available data have a huge number of speakers, languages and environmental variations, the unavailability of appropriate annotations (mostly done with automatic algorithms) poses a challenge for an in-depth analysis~\cite{khosravani2017plda}. The current challenge was planned with aim of resolving the above-mentioned issues by inviting the community to work on the development of the language and sensor invariant speaker representation.

This work considers the conversation recordings of the IITG-MV phase-I dataset. The dataset is divided into four parts, viz. (1) Development, (2) Enrollment, (3) Public, and (4) Private test set.  The development set consists of speech utterances from $50$ speakers recorded with all $5$ sensors and in $13$ languages. The enrollment set consists of utterances from the remaining $50$ speakers, spoken in English language  and through a headset microphone. The public test set consists of utterances from the $50$ enrolled speaker in both matched and mismatched sensors and languages. The private test set only consists of cross-lingual and sensor utterances. Along with releasing the dataset, the challenge was offered in the form of two sub-tasks, (1) constrained and (2) unconstrained. The constrained sub-task restricts the participants to use only the provided data. On the other hand, no such restrictions are there in the unconstrained sub-task. The aim of the constrained sub-task here was to encourage the community to develop the SV with limited training data. Conversely, the aim of the unconstrained sub-task was to observe the performance of SV technologies developed with a sufficient amount of training data. A baseline system implemented with X-vector framework for both constrained and unconstrained sub-tasks was made available to the participants during the challenge (available at \href{https://github.com/jagabandhumishra/I-MSV-Baseline} {\url{https://github.com/jagabandhumishra/I-MSV-Baseline}}). The performance of the baseline in public test data on both the sub-tasks were $9.32\%$ and $8.15\%$, respectively.

The rest of the paper is organized as follows: the challenge rules are described in section~\ref{2}. The detailed description of the data preparation is described in section~\ref{3}. Section~\ref{4} reports the procedure of baseline system development and the performance measure used. A brief description of the top five systems along with their performance are  described in section~\ref{5}. Finally, the summary and future directions are reported in section~\ref{6}.

\section{Challenge Rules}
\label{2}
As mentioned in the earlier section, the challenge consisted of two sub-tasks, viz. (1) constrained SV and (2) unconstrained SV.
\begin{itemize}
\item \textbf{Constrained SV}: Participants were not allowed to use speech data other than the speech data released as a part of the constrained SV challenge for the development of the SV system.

\item \textbf{Unconstrained SV}: Participants were free to use any publicly available speech data in addition to the audio data released as a part of unconstrained SV.
\end{itemize}

The challenge was organized as a part of the $25^{th}$ edition of the O-COCOSDA-2022 conference along with the Asian-Multilingual Speaker Verification (A-MSV) track. The participants were asked for registration. Upon agreeing to the data usage licenses agreement, the download link of the development, enrollment, and public test sets were provided. Through a license agreement, the participant teams agreed that they could use the data only for research purposes. Moreover, the top five systems in both the sub-tasks would have to submit the source code of their systems and a detailed report. 

The public test set released during the time of registration had ground truth information. The purpose here was to tune the system parameter using the public test data. The participants were asked to upload their score files in a specific format on the challenge portal. The corresponding performance was evaluated by a back-end script and the results were uploaded to a online leader board. There was no constraint on uploading and evaluating the score files on the public test set. After around one month of the public test set release, the private test set was released without ground truth information. The participant teams were asked to submit their final results on the private test set within $24$ hours from the release of the private test set. A maximum of three successful attempts were allowed for each team for evaluating their system on the private test set.

\section{Data Preparation}
\label{3}
The IITG-MV speaker recognition dataset was recorded in four phases for dealing with various speaker recognition applications, viz. speaker identification, verification, and change detection, etc.~\cite{haris2012multivariability}. Among the four phases, the phase-I dataset is considered for this study. The IITG-MV-Phase-I dataset consists of recordings from $100$ speakers in reading and conversation mode. In both modes, each speaker has given their speech data in two sessions. The duration of each session is around $5-8$ minutes. In addition, each speaker has given their data in two languages, viz. English and favorite language. Favorite language mostly meant their mother tongue/native language and varied from person to person. Furthermore, all the speech utterances were recorded through five different sensors, viz. H01, M01, M02, D01 and T01. The details of the dataset can be found at~\cite{haris2012multivariability}. The utterances belonging to the conversation mode were only considered here. The total duration of the selected utterances is approximately $100$ hours. The selected utterances are named as the I-MSV dataset. Further, the I-MSV dataset is segregated into four parts, viz. development, enrollment, public test, and private test.

\subsection{Development set} 
This partition consists of recordings from $50$ speakers. The utterances from each speaker are available in two languages, with two sessions, and with five sensors. The approximate duration of the development set is $50$ hours.

\subsection{Enrollment set} 
This partition consists of recordings from $50$ speakers that are disjoint from the speakers used in the development set. The utterances belonging to both the sessions with the English language and the Headset (H01) sensor are used here. The first session utterances are completely used in this set. However, the utterances from the second session are segmented into two parts. Half of them are used in enrollment and the rest have been used in the public test set (to observe the performance in matched sensor and language conditions). The approximate duration of speech available for each speaker is $8-10$ minutes.

\subsection{Public test set} 
This set consists of the utterances from the second session recordings with three sensors and cross-languages along with the matched utterances. The second session utterances in the original IITG-MV-Phase-I dataset are segregated into two parts. Half of them are reserved for the preparation of the private test set. After that, each utterance is segmented into $10$, $30$, and $60$ second utterances. The segments are split into silence regions using the knowledge of Voice Activity Detection. The segmented files were made available to the participants as the public test set. The total number of utterances available in this partition is $5907$.

\subsection{Private test set} 
This set consists of the utterances from the second session recordings with four sensors and cross-languages. This partition does not consist of matched sensors and language utterances. The selected utterances are segmented into $10$s, $30$s, and $60$s second utterances and made available to the participants as the private test set. The total number of utterances available in this partition is $9521$. The partition consists of cross-language utterances from $10$ Indian languages.

\begin{table}[!t]
\centering

\caption{Baseline results on I-MSV dataset}
\label{baseline_perf}

\begin{tabular}{ 
|p{0.2\columnwidth}
|p{0.3\columnwidth}
|p{0.3\columnwidth}
|} 

\hline

& \multicolumn{2}{c|}{\textbf{EER} (\%)} \\
\cline{2-3}

\textbf{Model} & \textbf{Overall} & \textbf{Matched Sensor and Language} \\ 
\hline

I-vector & $13.72$ & $4.61$ \\
\hline

X-vector & $9.32$ & $2.40$ \\
\hline

X-vector (unconstrained) & $8.15$ & $0.82$ \\

\hline

\end{tabular}
\end{table}

\begin{table*}[!t]
\centering
\caption{Summary of top $5$ submissions to the challenge. FE:=\emph{Frontend}, LF:=\emph{Loss Function}, BE:=\emph{Backend}, C-SV:={Constrained-SV}, UC-SV:={Unconstrained-SV}.}
\label{submission_summary}

\begin{tabular}{
|p{0.05\linewidth}
|p{0.2\linewidth}
|p{0.22\linewidth}
|p{0.22\linewidth}
|p{0.08\linewidth}
|p{0.08\linewidth}
|} 

\hline

 & & & & \multicolumn{2}{c|}{\textbf{EER} (\%)} \\
\cline{5-6}

\textbf{Team} & 
\textbf{BE} & 
\textbf{LF} & 
\textbf{FE} & 
\textbf{C-SV} & 
\textbf{UC-SV} \\ 
\hline

team0 & Rawnet3 & Training: triplet margin loss; Fine-tuning: AAM Loss + K-Subcenter loss + Inter-topK loss & Cosine similarity & -- & $0.26$ \\
\hline

team1 & ResNet with SE attention & Softmax + Angular Prototypical Loss & Model scoring (DNN, Random Forest and Gradient Boosting Trees) & -- & $0.36$ \\
\hline

team2 & ECAPA-TDNN + SE-ResNet blocks & Weight Transfer loss + AAM-Softmax loss + L2 loss & Cosine similarity & $2.12$ & $0.63$ \\
\hline

team3 & ECAPA-TDNN SE-ResNet blocks & AAM Loss & Cosine similarity & $2.77$ & $2.70$ \\
\hline

team4 & ECAPA-TDNN + SE-ResNet blocks & Large
Margin Cosine Loss & PLDA & 2.97 & $2.97$ \\

\hline

\end{tabular}
\end{table*}

\section{Performance Measures and Baselines}
\label{4}
This challenge employs the Equal Error Rate (EER) measure to compare the performances of the different submissions with the baseline results. This section briefly describes the method of computing the EER measure and reports the baseline results on the I-MSV dataset. Let, $N_{P}$ and $N_{N}$ be the number of positive and negative test samples in the data, respectively. The number of samples out of a total of $N_{P}$ positive samples predicted as positive are termed as True Positives (TP). On the other hand, the number of samples out of a total of $N_{N}$ negative samples correctly predicted as negative are termed as True Negatives (TN). Incorrectly predicted positive and negative samples are termed as False Positives (FP) and False Negatives (FN), respectively. The prediction of a test sample as positive or negative is based on a pre-determined threshold $\tau$ which may be varied. The total number of TP, TN, FP, and FN for the whole test data can be used to compute two measures, viz., False Acceptance Rate (FAR) and False Rejection Rate (FRR). The FAR can be defined using eq.~\ref{far}.

\begin{equation}
\label{far}
\text{FAR}=\dfrac{FP}{FP+TN} 
\end{equation}

\noindent Similarly, the FRR can be defined as in eq.~\ref{frr}.

\begin{equation}
\label{frr}
\text{FRR}=\dfrac{FN}{TP+FN} 
\end{equation}

\noindent When $\tau$ is varied, different values of FAR and FRR can be obtained. Among all the different $\tau$ used, a specific threshold $\tau_{equal}$ can be identified which provides equal (or almost equal) values of FAR and FRR. The EER measure is computed as the mean of FAR and FRR at $\tau_{equal}$ (eq.~\ref{eer}).

\begin{equation}
\label{eer}
\text{EER}=\dfrac{1}{2} \left(FAR+FRR\right)
\end{equation}

\noindent where, $\mid \text{FAR}-\text{FRR}\mid \to 0$.

The challenge organizers provided results on the I-MSV dataset using Kaldi based I-vector and X-vector systems as a baseline for comparison. The baseline performances are reported in Table~\ref{baseline_perf}.

\begin{figure*}[!t]
\centerline{
\includegraphics[width=0.7\linewidth]{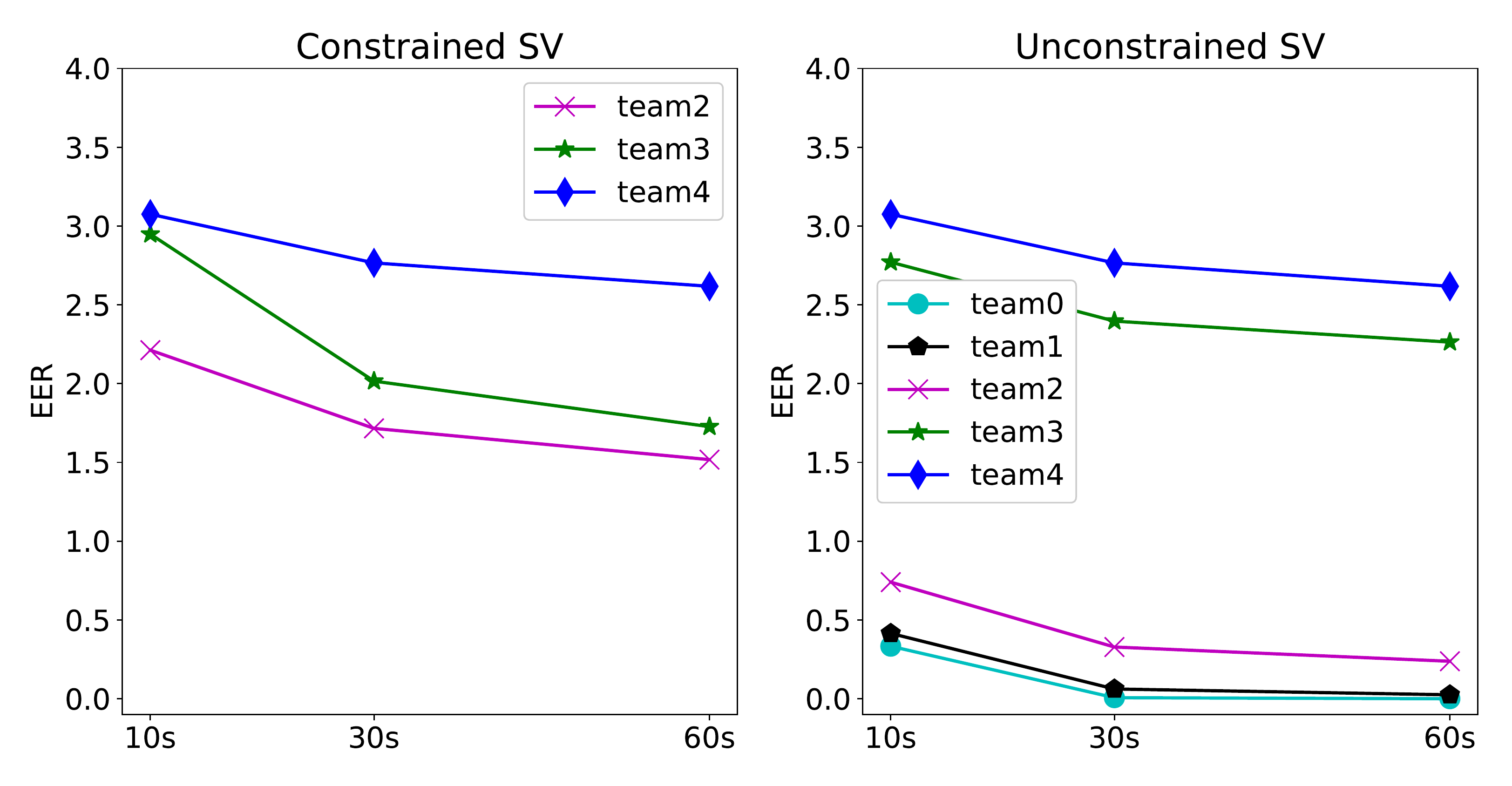}
}
\centerline{
(a)
}
\centerline{
\includegraphics[width=0.7\linewidth]{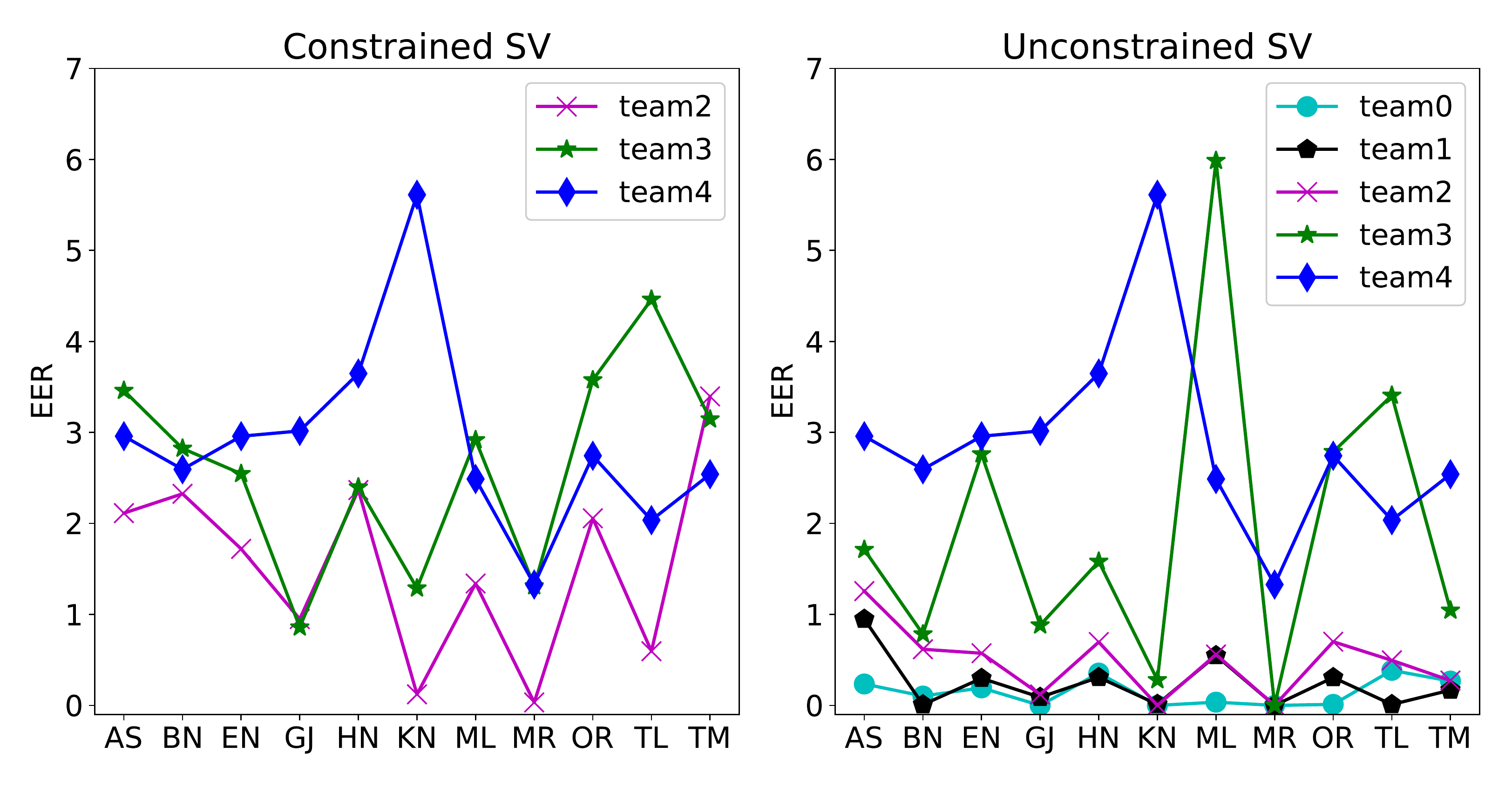}
}
\centerline{
(b)
}
\centerline{
\includegraphics[width=0.7\linewidth]{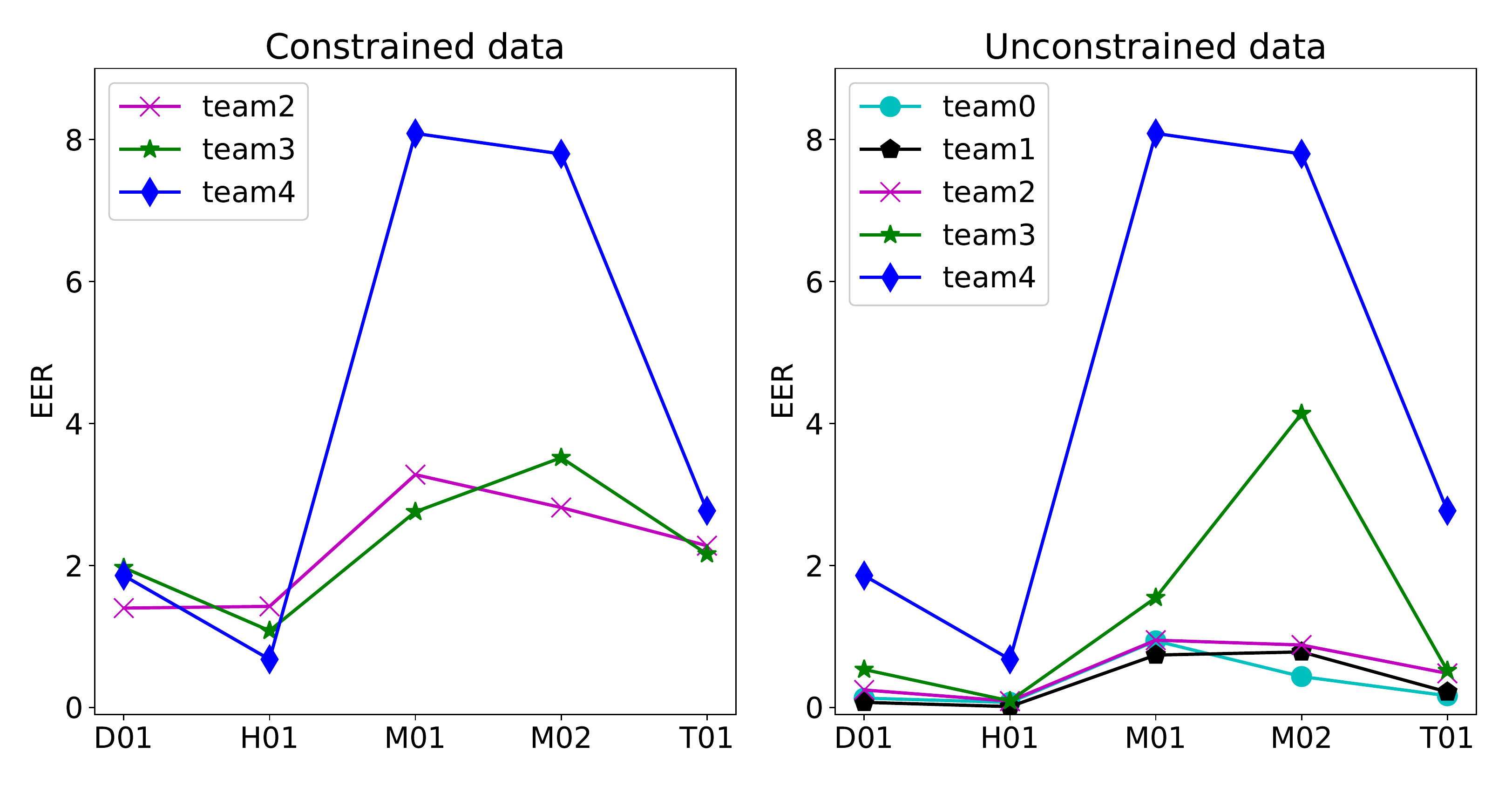}
}
\centerline{
(c)
}
\caption{Illustrating the effect of (a) different duration, (b) different languages, and (c) different sensors on the performance of submitted systems.}
\label{fig:duration_language_sensor_effect}
\end{figure*}

\section{Systems and Results}
\label{5}

A total of $25$ teams registered for the I-MSV 2022 challenge. Among these, $10$ teams submitted their results for the public test set evaluation. For the private test set evaluation, a total of $6$ teams submitted their results and systems. The best $5$ participating systems are summarised in the next paragraph. Table~\ref{submission_summary} lists a brief summary of the top $5$ systems. 

The submission of \emph{team0} obtained the best EER  of $0.26$ on the private test set using unconstrained training data. The best system of \emph{team0} used the Rawnet3 architecture~\cite{jung2022raw} as their front-end system. They initially trained the model with a Triplet Margin loss~\cite{BMVC2016_119}. Subsequently, they fine-tuned their model with a combination of Adaptive Angular Margin (AAM) K-Subcenter loss~\cite{deng2019arcface} and Inter-TopK loss~\cite{zhao2022multi}. They performed the backend scoring using the cosine-similarity measure and used adaptive score normalization. 

The second best EER of $0.36$ using unconstrained data was obtained by \emph{team1}. They used the ResNet-34 architecture proposed in~\cite{heo2020clova} with Attentive Statistics Pooling~\cite{okabe18_interspeech} for their front-end. They trained the model using a combination of vanilla Softmax loss and Angular Prototypical loss~\cite{chung20b_interspeech}. They also proposed a two-layer model scoring system composed of Fully-Connected Feed-Forward layers, Random Forests and Gradient Boosting Trees. 

The EER obtained by \emph{team2} on the constrained data scenario was $2.12$. They achieved an EER of $0.63$ using unconstrained training data. They used combination of ECAPA-TDNN~\cite{desplanques20_interspeech} and ResNet-34~\cite{heo2020clova} with Squeeze-and-Excitation (SE) attention as front-end models to obtain the best results in the constrained data scenario. However, only the ResNet-34-SE network provided the best performance in the unconstrained scenario. For the unconstrained scenario, they fine-tuned the backbone model using a combination of Weight-Transfer loss~\cite{zhang2022npu}, AAM-Softmax loss and $L_{2}$ loss. The backend scoring was performed using cosine similarity measure. 

The \emph{team3} obtained an EER of $2.77$ in the constrained scenario and and EER of $2.70$ in the unconstrained scenario. They used a similar front-end system as that of \emph{team2} and trained it using the AAM loss. They also performed the backend scoring using cosine similarity.

The EER obtained by \emph{team4} in the unconstrained scenario was $2.97$. They also employed a similar front-end architecture as that of \emph{team2} and used the Large Margin Cosine loss for training. They performed the backend scoring using Probabilistic Linear Discriminant Analysis (PLDA)~\cite{jiang12_interspeech}.

\section{Summary and Discussion}
\label{6}

The results obtained by the submitted systems can be summarised along the following broad directions. First, use of unconstrained training data is hugely beneficial in performing SV in low-resource scenario like the current challenge. Second, automatic feature learning and end-to-end models can learn highly discriminating features. Third, the choice of loss function for the front-end system has a huge impact on the obtained performance of similar architectures. Fourth, simple backend scoring like cosine similarity might be enough if the learnt speaker embedding are highly discriminating. Fifth, longer utterances (Fig.~\ref{fig:duration_language_sensor_effect}(a)) are more helpful in identifying the speakers. Sixth, change in language (Fig.~\ref{fig:duration_language_sensor_effect}(b)) degrades the SV performance. However, it might also be noted that such an observation may also be the result of imbalance in the number of utterances for the different languages in the I-MSV dataset. Seventh, the change in sensor (Fig.~\ref{fig:duration_language_sensor_effect}(a)) has a huge impact on the performance of SV systems. More specifically, SV systems fare poorly when presented with telephone channel recordings. In future, better SV systems may be developed by taking into consideration the observations made in this challenge.

\section*{Acknowledgments}
The authors like to acknowledge Ministry of Electronics and Information Technology (MeitY), Govt. of India, for supporting us through "Bhashini: Speech technologies in Indian languages" project. We are  also grateful to K. T. Deepak, Rajib Sharma and team (IIIT Dharwad, Karnataka), S. R. Nirmala, S. S. Chikkamath and team (KLETech, Hubballi, Karnataka), Debadatta Pati, Madhusudan Singh and team (NIT Nagaland, Nagaland), Joyanta Basu, Soma Khan and team (CDAC Kolkata, WB), Akhilesh Kumar Dubey, Govind Menon and team (KLU Vijayawada, AP), Gayadhar Pradhan, Jyoti Prakash Singh and team (NIT Patna, Bihar), and S. R. M. Prasanna, Gayathri A. and team (IIT Dharwad, Karnataka) for their help and cooperation in successfully organizing this challenge.

\bibliography{acl2020}
\bibliographystyle{ieeetr}

\end{document}